# Analysis of Test Efficiency during Software Development Process

T.R. Gopalakrishnan Nair
Research and Industry Incubation Center
Dayananda Sagar Institutions
Bangalore, India
trgnair@ieee.org

V. Suma
Research and Industry Incubation Center
Dayananda Sagar Institutions
Bangalore, India
sumavdsce@gmail.com

Pranesh Kumar Tiwari
Research and Industry Incubation Center
Dayananda Sagar Institutions
Bangalore, India
praneshtiwari@gmail.com

*Abstract*— One of the prerequisites of any organization is an unvarying sustainability in the dynamic and competitive industrial environment. Development of high quality software is therefore an inevitable constraint of any software industry. Defect management being one of the highly influencing factors for the production of high quality software, it is obligatory for the software organizations to orient them towards effective defect management. Since, the time of software evolution, testing is deemed a promising technique of defect management in all IT industries. This paper provides an empirical investigation of several projects through a case study comprising of four software companies having various production capabilities. The aim of this investigation is to analyze the efficiency of test team during software development process. The study indicates very low-test efficiency at requirements analysis phase and even lesser test efficiency at design phase of software development. Subsequently, the study calls for a strong need to improve testing approaches using techniques such as dynamic testing of design solutions in lieu of static testing of design document. Dynamic testing techniques enhance the ability of detection and elimination of design flaws right at the inception phase and thereby reduce the cost and time of rework. It further improves productivity, quality and sustainability of software industry.

## I. INTRODUCTION

The dynamic and competitive tendency of industrial market demands persistent effort of any organizations to develop high quality software. Developing high quality software within scheduled time, cost and resources is therefore one of the major concern of any software industry. Quality can be conceived by two dimensions namely through process quality and through people quality. Accordingly, software quality is defined as

$$\text{Software Quality} = \sum_{i=1}^{n} f(\text{Process Quality} + \text{People Quality}) \quad (1)$$

where i= requirements phase and n = maintenance phase of software development process.

Any activity associated with quality dimensions is continuous in nature and hence the impact of these activities is coupled to subsequent quality decisions made during the development process. Since, people drive the process, the quality of software development process is controlled by the quality level of the people. Consequently, it is vital to carry the software development activities with team consisting of good skill set.

It is indubitably known that testing is the last opportunity for the software industries to develop high quality software. An effective involvement of testing team during software development process is therefore a promising strategy for the organization to get an intended quality in software. Despite the existence of software test life cycle (STLC) in most of the matured software industries, current trend in industry typically deems the significance of testing as a critical activity during the code testing stage of development.

Connotation of testing at requirements analysis and design phase is less emphasized because of which defects propagates and amplifies at later stages of software development. Further, cost and time to fix the defect either at later stages of software development or for the rework purpose is very expensive. Therefore, it is very vital for the software industries to comprehend the significance of defect management techniques in order to develop high quality software products [1][2][3].

This paper brings out an analysis of test team at three major phases of software development namely requirements analysis phase, design and implementation phase through a case study comprising of two leading service–based and two leading product-based software industries. An investigation from the case study infers that efficiency of test team at requirements phase and design phase is low when compared to the test efficiency at the implementation phase of development. Subsequently, the study focused on analyzing the test effort at the design phase across the companies under study. The paper further suggests modern approaches of testing in order to improve the efficiency of test team at the design phase of development process.

## II. BACKGROUND WORK

Authors in [4] express that software organizations can improve their quality within estimated cost and schedule with good understanding of effort distribution of quality assurance activities such as reviews, process audit and testing. They feel that an awareness of effort distribution leads towards process improvement. Therefore, success of an organization depends upon well-structured management systems, quality management approach and on the methodologies adopted for continuous improvement [5]. Therefore, it is significant for the organizations to produce quality products to provide responsive services and to enhance the customer's value for the continued existence in the competitive market [6].

Authors in [7] state that quality is free while non-quality things cost. Hence, they suggest that the best practice of identification and fixing of process defects enable one to achieve the product quality. Defect detection and elimination is therefore one of the major challenges in the realization of high quality products [1] [2] [3].

An author in [8] expresses the lack of research in testing domain and states the existence of test gap due to the inconsistency of research and practices. He recommends research in testing domain to progress with real time problems in lieu of hypothetical problems.

Authors of [9] state that testing at design phase is very important as design errors and defects are harmful to the application. They suggest the implementation of architectural testing in addition to coding and unit testing. According to report in [10], the current challenge of software industry is to manage the threats faced during operations such as accidental design or implementation errors. Best practice of effective testing at each phase enables the organization to overcome such threats and henceforth assures the quality of the product. Therefore, author in [11] suggests the testers to interact with designers in order to resolve testability issues and to develop integration test plans based on design documents. Authors of [12] recommend a systematic testing, which demands an effort to match the software design and its corresponding test cases during integration testing in order to enhance quality level of software testing. However, authors of [13] feel that with change in product design, design test becomes outmoded in the V model of software testing approach. Therefore, they suggest design inspections and code reviews to improve the quality of the product with less time in lieu of long hours of pre-writing tests.

Testing is an indispensable activity of software development process. Author of [14] states that the project team in an organization uses the information provided by testing team in order to make improvements in the project and in the system. Consequently, the process of testing influences and is in turn influenced by the project and the system. This paper henceforth brings in awareness of significance of testing and the process of testing to develop high quality software product with focal point of testing at design phase.

### III. RESEARCH DESIGN

Owing to the sphere of influence of software across the vicinity of human life, the aim of this investigation was to analyze the test efficiency during software development process. This paper however focuses on the test efficiency during design phase through a case study comprising of four leading software industries. Company A and Company B are leading CMMI 5 certified service-based software industries while Company C and Company D are leading CMMI 5 certified product-based software industries. Data collection is through data centers and quality assurance departments of the aforementioned industries.

In order to overcome the challenge of development of wide spectrum of projects, this investigation considers projects developed from the year 2000 onwards to 2011. These projects are development type of projects, which includes maintenance of information of the entire organization such as inventory information, manufacturing plant information etc. Executives of the organization are therefore able to make internal decision towards the smooth functioning of the organization. These projects are developed in Windows XP Operating System using Visual Studio 2003, 2005 and 2008. The empirical projects under study are developed using C++, VC++ and MFC language. Several projects were investigated in this context and this paper presents a collection of 20 projects by considering a sample of 5 projects from each company under study.

Hypothesis: The projects under study are developed using aforementioned programming language and within the above specified development environment.

Threat to hypothesis: The design efficiency pattern observed may not be applicable to legacy projects, innovative projects and for projects with low technical risks.

### IV. CASE STUDY

The companies under study are a CMMI 5 certified software industry, which functions with business application projects. This investigation focuses on projects developed with stand-alone applications meant for Information System Management. The companies follow review, inspection and testing strategies as a mode of quality assurance techniques.

Table I illustrates a sample of 5 projects collected from Company A. They depict the defect capturing capabilities of the test team within the constraints of time, number of testers involved along with their average experience level in the project at three major phases of software development namely requirements analysis phase, design phase and implementation phase. Similarly, Table II, Table III and Table IV represent the aforementioned data information of Company B, Company C and Company D respectively. It may be noted that the project development time in the above specified tables is measured in person hours where

Project development time = (8 hours of work per day) × (number of software personnel) × (number of months required)

(2)

Figure 1. through Figure 3. depicts the comparative test efficiency analysis of the four companies under study at the three major phases of software development. Figures 4. through Figure 7. depict test effort analysis during design phase of the four companies.

### V. INFERENCES AND DISCUSSIONS

It is apparent from the Figures 1. through Figure 3. that despite of process maturity of the four software companies, the test efficiency at requirements analysis and design phase is considerably very low. However, it is worth to note that design phase is one of the vital phases of software development. This awareness focused the investigation towards an analysis of the test efficiency of each company at design phase. Figures 4. through Figures 7. indicate the comparative test efficiency of each company against the sampled projects at design phase of software development in compliance with their test effort.

The figures indicate that defect capturing ability of test team within the specified constraints of testing time, number of testers and their average test experience level is extremely low in all the companies under investigation. Rationale for this poor defect capturing ability of test team across the companies under study is due to factors such as less involvement of testers for testing design solutions, lack of domain knowledge, lack of design skills, insufficient testing time etc. This calls for immediate actions to improve the defect capturing strategies of the test team at design phase.

Despite the differences observed in the development process and production capabilities of the software organizations, there exists a generic design development approach. Senior architect provides a design solution to the specified requirements, which is either self reviewed or inspected to identify static design imperfections. Nevertheless, architects perform knowledge transfer (KT) of design solutions to the test team, mutual interactions between design and test team is very less. Habitually, test team perform static test of the design document to detect the defects while the code developers perform dynamic execution of the design solution. Consequently, defect detection rate is low as designers cannot visualize the design flaws until the design is implemented. This leads towards accumulation of design defects and overhead of rework to fix these defects in terms of required cost, time, and resources and on the desired level of quality in the software to be developed.

This paper hence emphasize on modern approach of test strategies, which includes involvement of test team during design phase. This can be achieved in software industry using dynamic execution of design solutions through techniques such as prototype to investigate the design flaws at the design phase. In this mode of testing, the design solutions, which are, currently self-reviewed by the design team is now recommended to undergo testing by the test team in presence of design team in order to identify and rectify the design defects. However, it is further suggested to prototype only the critical features of the application. The study further opens the software community to choose their desired strategies, which can serve the aforementioned purpose of dynamic execution of design solutions in collaboration between design and test team. The aim of dynamic execution of design is to detect the design defects at the phase. It enables the testers to involve in the design phase, which in turn enhances their defect capturing skill since design team and test team, work concurrently. Defect identification by the test team at the phase reduces the inherent nature of defects to amplify and propagate into later phases of development. It improves quality, productivity and assured sustainability of the company in the competitive industrial atmosphere. Nevertheless, optimality of modern techniques for dynamic execution of design solutions by design team as a group effort by test and design team depends on time, cost, and other such associated resource constraints.

VI. CONCLUSISONS

Development of quality software product is the rudimentary need of any software industry to sustain them in the dynamic and competitive industrial atmosphere. Quality is a continuous process and not a state, which is dependent on process and people. Effective defect management is deemed one of the significant and influencing parameters of quality.

Testing is a well-established defect management technique since the evolution of software. Despite the advancement in the process of software development, efficiency of test team in effective defect detection at the major phases of software development namely requirements analysis phase, design phase and implementation phase does not have consistent results.

This paper provides an empirical investigation of several projects through a case study comprising of four software companies. It includes two leading service-based software industries and two leading product-based software industries. The aim of this investigation is to analyze the efficiency of test team in effective defect detection across the companies having various production capabilities. The case study indicates a low defect capturing ability of the test team at requirements analyses phase and even lower test efficiency at design phase in comparison with implementation phase of software development. A further analysis of test effort in the four companies at the design phase indicates lack of effective involvement of test team resulting in poor test performance.

It is suggested to implement modern approaches of testing to detect design flaws right at the design phase using techniques such as dynamic execution of design solutions in lie of static testing of design documents. Implementation of modern approaches of testing with involvement of test team in association with design team enhances the skill set of test team and thereby their efficiency level of defect detection. Consequently, design imperfections are resolved at the design phase, which reduces cost, time and resources required for rework. This further leads towards increased productivity, quality and sustainability of the software industry.

ACKNOWLEDGMENT

Authors would like to thank all the industry people who extended their valuable support and help in compliance within the framework of non-disclosure agreement.

TABLE I  TEST ANALYSIS OF COMPANY A

| | Project | p1 | p2 | p3 | p4 | p5 |
|---|---|---|---|---|---|---|
| | Project Development Time (*) | 22000 | 27500 | 33000 | 55000 | 60500 |
| | No of Lines of code (**) | 55000 | 66000 | 77000 | 121000 | 132000 |
| Req Phase | Requirements Time (*) | 2200 | 2200 | 3300 | 5500 | 5500 |
| | Testing Time (*) | 220 | 220 | 220 | 660 | 660 |
| | Number of testers | 3 | 3 | 3 | 4 | 5 |
| | Experience level (***) | 4 | 4 | 4 | 6 | 6 |
| | % Estimated defects | 16.2 | 13.6 | 13.5 | 14.8 | 9.9 |
| | % Defects captured | 6.6 | 7.7 | 8.3 | 5.9 | 6.8 |
| | % Defects un-captured | 9.6 | 5.9 | 5.2 | 8.9 | 3.1 |
| | % Bad fixes | 1.9 | 1 | 2 | 1.9 | 1 |
| | Total defects un-captured | 11.5 | 6.9 | 7.2 | 10.8 | 4.1 |
| Des phase | Design Time (*) | 4400 | 4950 | 5500 | 15400 | 12100 |
| | Testing Time (*) | 550 | 660 | 715 | 2200 | 2420 |
| | Number of testers | 3 | 3 | 3 | 4 | 5 |
| | Experience level (***) | 4 | 4 | 4 | 6 | 6 |
| | % Estimated defects | 11.7 | 9.6 | 12.8 | 9.3 | 7.6 |
| | % Defects captured | 4.4 | 3.2 | 3.9 | 5.1 | 5.5 |
| | % Defects t un-captured | 7.3 | 6.4 | 8.9 | 4.2 | 2.1 |
| | % Bad fixes | 0 | 0 | 0 | 0 | 1.1 |
| | Total defects un-captured | 7.3 | 5.2 | 6.3 | 3.3 | 3.4 |
| Imp Phase | Implementation time | 15400 | 20350 | 24200 | 34100 | 42900 |
| | Testing Time (*) | 3750 | 4000 | 4250 | 11375 | 11875 |
| | Number of testers | 3 | 3 | 3 | 4 | 5 |
| | Experience level (***) | 3 | 3 | 3 | 5 | 5 |
| | % Estimated defects | 44.75 | 35.15 | 36.05 | 25.05 | 20.8 |
| | % Defects captured | 45 | 38.75 | 37.5 | 27.5 | 19.9 |
| | % Bad fixes | 3.3 | 2.75 | 3.05 | 3.45 | 2.95 |

Req-Requirements analysis phase; Des-Design phase; Imp-Implementation phase ;
(*) - Measured in person hours; (**) – Kilo lines of code; (***) – measured in years

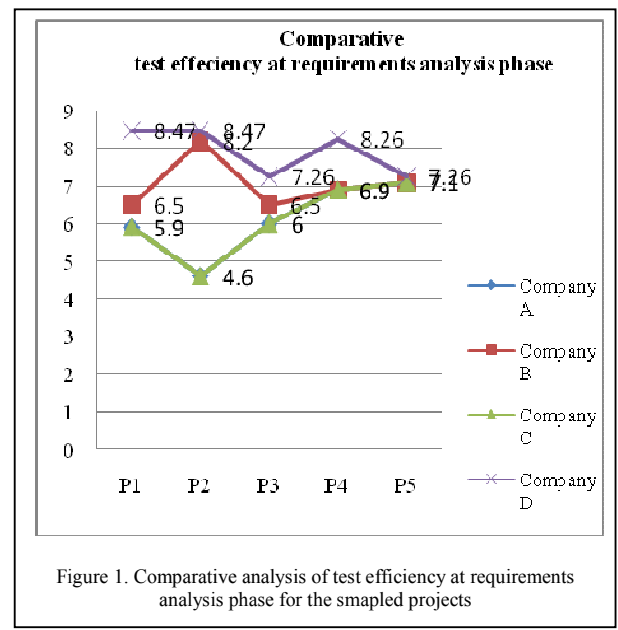

Figure 1. Comparative analysis of test efficiency at requirements analysis phase for the smapled projects

TABLE II   TEST ANALYSIS OF COMPANY B

| Project | | p1 | p2 | p3 | p4 | p5 |
|---|---|---|---|---|---|---|
| Project Development Time (*) | | 19800 | 22000 | 25300 | 49500 | 52800 |
| No of Lines of code (**) | | 55000 | 55000 | 55000 | 110000 | 110000 |
| Req Phase | Requirements Time (*) | 1980 | 2090 | 2530 | 1980 | 5720 |
| | Testing Time (*) | 220 | 231 | 264 | 660 | 737 |
| | Number of testers | 3 | 3 | 3 | 4 | 4 |
| | Experience level (***) | 4 | 4 | 4 | 6 | 6 |
| | % Estimated defects | 17.8 | 15.5 | 12.3 | 14 | 12.5 |
| | % Defects captured | 6.5 | 8.2 | 6.5 | 6.9 | 7.1 |
| | % Defects un-captured | 11.3 | 7.3 | 5.8 | 7.1 | 5.4 |
| | % Bad fixes | 1.1 | 2.1 | 1.2 | 1.8 | 1.5 |
| | Total defects un-captured | 12.4 | 9.4 | 7 | 8.9 | 6.9 |
| Des Phase | Design Time (*) | 3850 | 4070 | 4290 | 9900 | 10450 |
| | Testing Time (*) | 550 | 572 | 627 | 1760 | 2090 |
| | Number of testers | 3 | 3 | 3 | 4 | 4 |
| | Experience level (***) | 4 | 4 | 4 | 6 | 6 |
| | % Estimated defects | 11.8 | 9.6 | 9.8 | 8.6 | 7.7 |
| | % Defects captured | 6.3 | 3.9 | 4 | 4.3 | 5.2 |
| | % Defects un-captured | 5.5 | 5.7 | 5.8 | 4.3 | 2.5 |
| | % Bad fixes | 0 | 0 | 0 | 0 | 1.1 |
| | Total defects un-captured | 5.5 | 5.7 | 5.8 | 4.3 | 3.6 |
| Imp Phase | Implementation Time (*) | 13970 | 15840 | 18480 | 37620 | 36630 |
| | Testing Time (*) | 3750 | 3750 | 3750 | 11250 | 11250 |
| | Number of testers | 3 | 3 | 3 | 4 | 4 |
| | Experience level (***) | 3 | 3 | 3 | 5 | 5 |
| | % Estimated defects | 39.15 | 35.1 | 34.05 | 18.2 | 19.25 |
| | % Defects captured | 45 | 38.75 | 31.5 | 27.5 | 26.25 |
| | % Bad fixes | 2.75 | 3 | 3.75 | 2.6 | 3.9 |

Req-Requirements analysis phase; Des-Design phase; Imp-Implementation phase ;
(*) - Measured in person hours; (**) – Kilo lines of code; (***) – measured in years

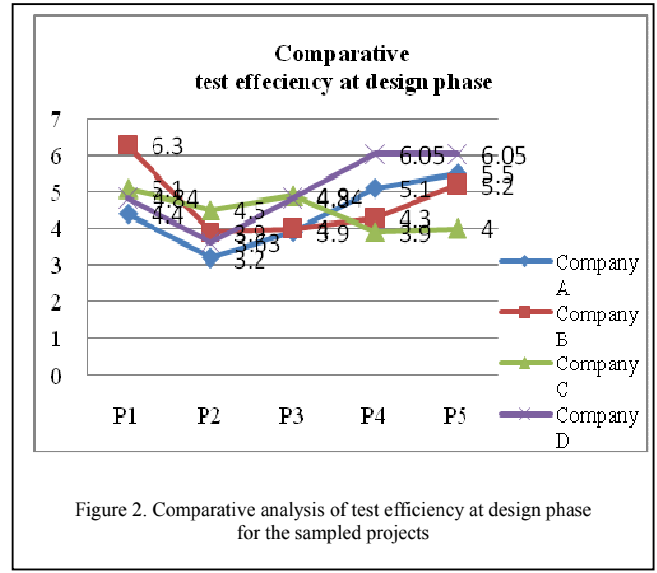

Figure 2. Comparative analysis of test efficiency at design phase for the sampled projects

TABLE III   TEST ANALYSIS OF COMPANY C

| Project | | p1 | p2 | p3 | p4 | p5 |
|---|---|---|---|---|---|---|
| Project Development Time (*) | | 82500 | 88000 | 93500 | 99000 | 110000 |
| No of Lines of code (**) | | 110000 | 110000 | 110000 | 121000 | 121000 |
| Req Phase | Requirements Time (*) | 8250 | 8800 | 9350 | 9900 | 11000 |
| | Testing Time (*) | 1100 | 1320 | 1650 | 2310 | 2200 |
| | Number of testers | 2 | 2 | 2 | 3 | 3 |
| | Experience level (***) | 6 | 6 | 6 | 6 | 6 |
| | % Estimated defects | 13.4 | 14.3 | 10.8 | 12 | 11.9 |
| | % Defects captured | 5.9 | 4.6 | 6 | 6.9 | 7.1 |
| | % Defects un-captured | 7.5 | 9.7 | 4.8 | 5.1 | 4.8 |
| | % Bad fixes | 2 | 1.1 | 1.3 | 2.1 | 1 |
| | Total defects un-captured | 9.5 | 10.8 | 6.1 | 7.2 | 5.8 |
| Des Phase | Design Time (*) | 16500 | 19800 | 22000 | 23100 | 28600 |
| | Testing Time (*) | 2640 | 2970 | 3300 | 3520 | 3740 |
| | Number of testers | 2 | 2 | 2 | 3 | 3 |
| | Experience level (***) | 6 | 6 | 6 | 6 | 6 |
| | % Estimated defects | 8.4 | 8.7 | 7.1 | 9.8 | 6 |
| | % Defects captured | 5.1 | 4.5 | 4.9 | 3.9 | 4 |
| | % Defects un-captured | 3.3 | 4.2 | 2.2 | 5.9 | 2 |
| | % Bad fixes | 1.1 | 1.1 | 1.1 | 1.1 | 1.1 |
| | Total defects un-captured | 4.4 | 5.3 | 3.3 | 7 | 3.1 |
| Imp Phase | Implementation time | 57750 | 59400 | 62150 | 66000 | 70400 |
| | Testing Time (*) | 18750 | 22500 | 23750 | 30000 | 36250 |
| | Number of testers | 3 | 3 | 3 | 3 | 3 |
| | Experience level (***) | 5 | 5 | 5 | 5 | 5 |
| | % Estimated defects | 28.4 | 27.9 | 20.15 | 27.35 | 21.9 |
| | % Defects captured | 35 | 32.3 | 28.6 | 30.1 | 23.3 |
| | % Bad fixes | 2.6 | 3.2 | 3.6 | 2.8 | 3.45 |

Req-Requirements analysis phase; Des-Design phase; Imp-Implementation phase ;
(*) - Measured in person hours; (**) – Kilo lines of code; (***) – measured in years

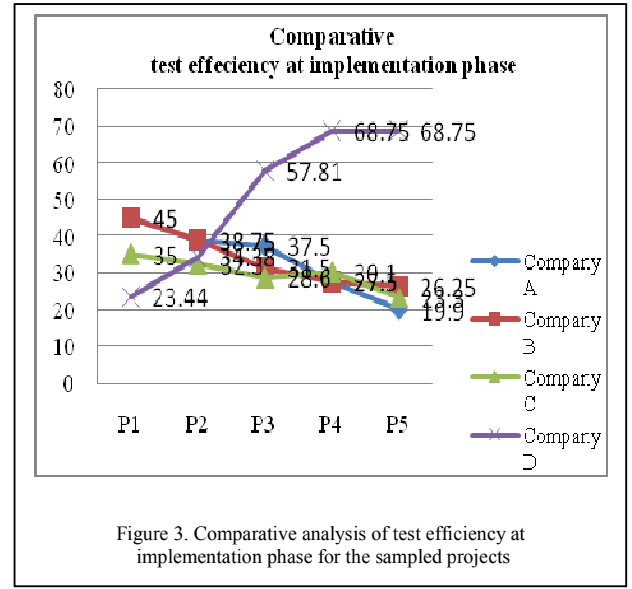

Figure 3. Comparative analysis of test efficiency at implementation phase for the sampled projects

TABLE IV   TEST ANALYSIS OF COMPANY D

| Project | | p1 | p2 | p3 | p4 | p5 |
|---|---|---|---|---|---|---|
| Project Development Time (*) | | 30250 | 37830 | 58080 | 90750 | 108900 |
| No of Lines of code (**) | | 44000 | 51000 | 78000 | 121000 | 133100 |
| Req phase | Requirements Time (*) | 2420 | 2783 | 6292 | 9075 | 10890 |
| | Testing Time (*) | 242 | 290 | 811 | 1210 | 2541 |
| | Number of testers | 3 | 3 | 3 | 2 | 2 |
| | Experience level (***) | 4 | 4 | 6 | 6 | 6 |
| | % Estimated defects | 16.8 | 16.8 | 14.8 | 17.1 | 15.7 |
| | % Defects | 8.47 | 8.47 | 7.26 | 8.26 | 7.26 |
| | % Defects un-captured | 8.33 | 8.33 | 7.54 | 8.54 | 7.54 |
| | % Bad fixes | 1 | 2 | 1 | 2 | 0 |
| | Total defects un-captured | 9.33 | 10.33 | 8.54 | 10.54 | 9.54 |
| Des phase | Design Time (*) | 5445 | 4719 | 11495 | 18150 | 25410 |
| | Testing Time (*) | 726 | 689.7 | 2299 | 2904 | 3872 |
| | Number of testers | 3 | 3 | 3 | 2 | 2 |
| | Experience level (***) | 4 | 4 | 6 | 6 | 6 |
| | % Estimated defects | 8.06 | 9.27 | 6.27 | 7.48 | 8.69 |
| | % Defects captured | 4.84 | 3.63 | 4.84 | 6.05 | 6.05 |
| | % Defects un-captured | 3.22 | 5.64 | 1.43 | 1.43 | 2.64 |
| | % Bad fixes | 1.1 | 1.1 | 1.1 | 1.1 | 1.1 |
| | Total defects un-captured | 4.32 | 6.74 | 2.53 | 2.53 | 3.74 |
| Imp Phase | Implementation Time | 22385 | 30328 | 40293 | 63525 | 72600 |
| | Testing Time (*) | 5000 | 4688 | 14063 | 23438 | 37500 |
| | Number of testers | 3 | 3 | 3 | 3 | 3 |
| | Experience level (***) | 3 | 3 | 5 | 5 | 5 |
| | % Estimated defects | 16.88 | 20.55 | 5.41 | 12.91 | 12.55 |
| | % Defects captured | 23.44 | 34.38 | 57.81 | 68.75 | 68.75 |
| | % Bad fixes | 4.69 | 4.69 | 4.69 | 4.69 | 4.69 |

Req-Requirements analysis phase; Des-Design phase; Imp-Implementation phase ;
(*) - Measured in person hours; (**) – Kilo lines of code; (***) – measured in years

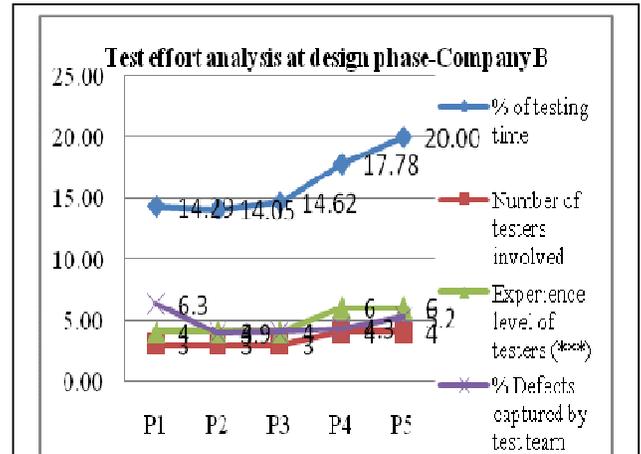

Figure 5. Test effort analysis of Company B at design phase
(***) years

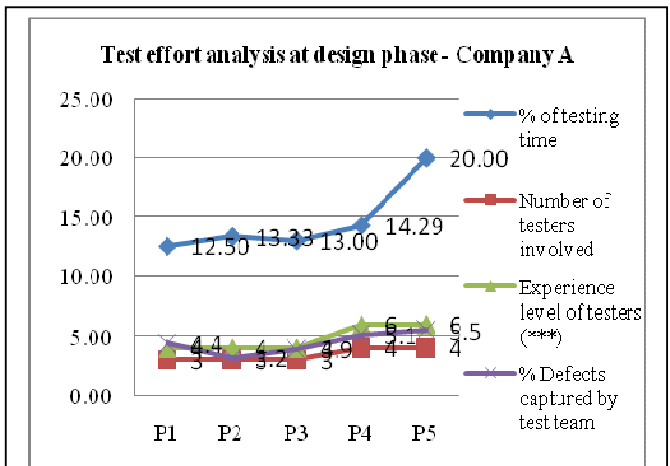

Figure 4. Test effort analysis of Company A at design phase
(***) years

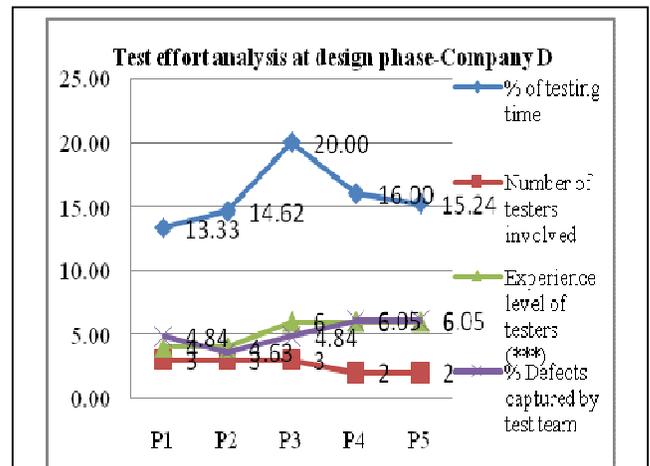

Figure 4. Test effort analysis of Company D at design phase